\documentstyle[12pt]{article}
\input{psfig.tex}

\font\elevenbf=cmbx10 scaled\magstep 1
\font\elevenrm=cmr10 scaled\magstep 1
 1
\def\be{\begin{equation}}
\def\ee{\end{equation}}
\def\bd{\begin{displaymath}}
\def\ed{\end{displaymath}}
\def\bea{\begin{eqnarray}}
\def\eea{\end{eqnarray}}
\def\nn{\nonumber}
\def\lb{\lbrack}
\def\rb{\rbrack}
\def\cz{\cos^2\hskip-2pt}
\def\cd{\cos^3\hskip-2pt}
\def\sz{\sin^2\hskip-2pt}
\def\sv{\sin^4\hskip-2pt}
\def\hpl{\hskip-2pt+\hskip-2pt}
\def\hmi{\hskip-2pt-\hskip-2pt}
\begin{document}
\hoffset=0.20truecm
\noindent
DESY 98-179 \hfill ISSN 0418-9833
\par\noindent
November 1998
\vskip40pt
\begin{center}{\elevenbf THE TOUSCHEK EFFECT \\ 
\vglue 3pt
IN STRONG FOCUSING STORAGE RINGS \\}
\vskip 70pt
{\elevenrm A. Piwinski \\}
\vglue 5.0cm
{\elevenbf Abstract}
\end{center}
\vglue 0.0cm
{\rightskip=4pc
\leftskip=4pc
\noindent
The lifetime of a stored beam due to the Touschek effect is 
calculated for arbitrary ratios of beam height to beam width. 
A variation of the beam envelopes is taken into account, i.\ e.\ 
the derivatives of the horizontal and vertical amplitude functions 
and dispersions are included. The calculation is done for arbitrary 
energies in the rest frame of the colliding particles.
\vglue 1.0cm}
\newpage
\pagestyle{plain}
\vglue 0.2cm
{\elevenbf\noindent 1.\ \ \,Introduction}
\vglue 0.4cm
\baselineskip=14pt
\elevenrm
\par
Coulomb scattering of charged particles in a stored beam causes
an exchange of energies between the transverse and longitudinal
motion. It changes, therefore, the betatron and synchrotron oscillation
coordinates of the colliding particles. One consequence is the
Touschek effect which  is the transformation of a small transverse
momentum into a large longitudinal momentum due to the
scattering. Then both scattered particles are lost,
one with too much and one with too little energy. The amplification 
of the momentum change is a relativistic effect so that the change of
the longitudinal momentum is increased by the Lorentz factor
$\gamma$.
\par
The Touschek effect is different from intra-beam scattering which is 
also caused by Coulomb scattering. The intra-beam scattering, however,
is a multiple scattering which leads to diffusion in all three
directions and, primarily, changes the beam dimensions. The
Touschek effect, on the other hand, is a single scattering effect
which leads to the immediate loss of the colliding particles.
Here only the energy transfer from the transverse to the longitudinal
direction plays a role
\par
The Touschek effect was observed for the first time in the small 
electron storage ring ADA $\lb1\rb$ and it turned out later that it 
is a serious problem in storage rings at low energies and in 
synchrotron light sources where it can reduce the lifetime considerably.
Theoretical investigations were made in $\lb1,2\rb$ for the
non-relativistic case, where the velocities of two colliding
particles are non-relativistic in their center of mass system. The theory
was then extended in $\lb3\rb$ to the ultra-relativistic case,
and in $\lb4\rb$ to arbitrary energies. Dispersion was included in
$\lb5\rb$. In all these calculations only the transfer of horizontal
momentum into longitudinal momentum was considered whereas the transfer
of the vertical momentum was neglected. In $\lb6\rb$ a 100\% coupling of
horizontal and vertical betatron oscillations, i.\ e.\ the special case of
a round beam, was considered for the non-relativistic case.
\par
The main aim of the following calculation is to take into account 
both the horizontal and the vertical betatron oscillations. In that case
one has a two-dimensional distribution of the transverse momenta which 
leads to a weaker dependence on the maximum stable energy deviation than a
one-dimensional distribution. This can increase the Touschek lifetime by a 
factor of two as compared to the flat beam approximation at energies 
around 1 GeV, but the correction can be larger at lower energies.
\par 
At the same time a variation of the beam envelopes is taken into    
account. The derivatives of the amplitude
functions $\beta_x$ and $\beta_z$ and of the dispersions $D_x$
and $D_z$ increase the angle between colliding particles. But the
derivatives of the amplitude functions play a role only at positions 
where the dispersion does not vanish, or to be exact, where
$\;\beta_{x,z}D^\prime_{x,z}-\beta^\prime_{x,z}D_{x,z}/2\;$ is
different from zero. 
In that case they can increase the loss rate and decrease the 
Touschek lifetime.
\par
We will use here the same method as for the calculation of the
intra-beam scattering $\lb7\rb$. Thus we will consider the
collision of two arbitrary particles in their center of mass system 
using the complete M{\o}ller scattering cross section, i.\ e.\ assuming
arbitrary energies in the center of mass system.
After transforming into the laboratory 
\newpage
\par\noindent
system we will calculate the
number of collisions with those momentum changes which lead
to the loss of both particles. When averaging  over the positions and
angles of the colliding particles we will assume  Gaussian 
distributions for all coordinates. 

\vskip 30pt
\par\noindent
{\bf 2.\ \ \,Change of the momenta of two colliding particles}
\vskip 10pt
\par
The momenta $\vec p_{1}$ and $\vec p_{2}$ of the two colliding particles 
before the collision are given in the laboratory coordinate system 
$\{s,x,z\}$ by the two vectors
\begin{eqnarray}
\vec p_{1,2} = \left(\begin{array}{c} p_{s1,2} \\ p_{x1,2} \\
 p_{z1,2} \end{array}\right)_{s,x,z}
\end{eqnarray}
where $s$ indicates the longitudinal, $x$ the horizontal, and $z$ 
the vertical direction.
We define a coordinate system with the axes $j$, $k$, and $l$
which are parallel to $\vec p_1+\vec p_2$, $\vec p_1\times \vec
p_2$, and $(\vec p_1+\vec p_2)\times (\vec p_1\times \vec p_2)$,
respectively. The two momenta then take the form
\be
\vec p_{1,2} \; = \; p_{1,2} \left(\begin{array}{c} \cos \chi_{1,2} \\ 0 \\
\pm \sin \chi_{1,2} \end{array}\right)_{j,k,l} \ee
where $\chi_1$ and $\chi_2$ are the angles between the vector 
$\vec p_1+\vec p_2$ and the vectors $\vec p_1$ and $\vec p_2$, 
respectively. They are given by the two relations
\be
\vec p_1\vec p_2\,=\,p_1p_2\cos(\chi_1\hpl\chi_2) \ee
and
\vskip-15pt
\be
p_1\sin\chi_1\,=\,p_2\sin\chi_2 \ee
\par
When we apply a Lorentz transformation parallel to $\vec p_1+\vec
p_2$ we obtain for the momenta $\vec{\bar p}_1$ and $\vec{\bar p}_2$
 in the c.\ o.\ m.\ system the representation
\be
\vec {\bar p}_{1,2} \;
= \; p_{1,2} \left( \begin{array}{c} \gamma_t(\cos \chi_{1,2}
-\beta_t/\beta_{1,2}) \\ 0 \\ \pm \sin \chi_{1,2} \end{array}\right)
_{\bar j,\bar k,\bar l} \ee
The transformed energies $\bar E_1$ and $\bar E_2$ are given by (see
App.\ A1)
\be
\bar E_{1,2}\,=\,\gamma_tE_{1,2}(1\hmi\beta_t\beta_{1,2}\cos
\chi_{1,2}) \,=\,(E_1\hpl E_2)/(2\gamma_t) \ee
where $\beta_t$ is the relative velocity of the c.\ o.\ m.\
system, $\gamma _t$ is the Lorentz factor of the transformation,
$\beta_{1,2}$ are the relative velocities of the two particles in the
laboratory system, and the bars denote all quantities in the c.\ o.\ m.\ 
system. $\beta_t$ is determined by the condition that the sum of the 
two momenta vanishes in the c.\ o.\ m.\ system and is given by:
\be
\beta_t\,=\,{c|\vec p_1+\vec p_2|\over E_1+E_2}
\,=\,{\beta_1\gamma_1\cos\chi_1+\beta_2\gamma_2\cos\chi_2
\over\gamma_1+\gamma_2 } \ee
\par
We may now assume that the quantities $\,(p_1\hmi p_2)^2/
p_{1,2}^2 $, $\,p_{x1,2}^2/p_{1,2}^2$, and $\,p_{z1,2}^2
/p_{1,2}^2$, and therefore also $\chi_{1,2}^2$, are small as
compared to one, and Eqs.(3) and (4) become, if we replace
$\,p_{s1,2}\,$ by $\,p_{1,2}\hmi(p_{x1,2}^2\hpl p_{z1,2}^2)/(2p_{1,2})$:
\bd
(p_{x1}/p_1\hmi p_{x2}/p_2)^2+(p_{z1}/p_1\hmi p_{z2}/p_2)^2\,=\,
(\chi_1+\chi_2)^2 \ed
and
\bd
p_1\chi_1\,=\,p_2\chi_2 \ed
~Finally one obtains with $\chi_1\approx\chi_2\approx\chi$
\be
4\chi^2\,=\,(p_{x1}-p_{x2})^2/p^2+(p_{z1}-p_{z2})^2/p^2 \ee
where $p$ is the mean value of all momenta in the bunch. Eqs.(5)
and (6) then simplify to (see App.\ A1)
\be
\vec {\bar p}_{1,2} \;
= \pm p \, \left(\begin{array}{c} \xi\sqrt{1+\gamma^2\chi^2}\,/2 \\ 0 \\
\chi \end{array}\right)_{\bar j,\bar k,\bar l}\, \ee
and
\be
\bar E_{1,2}\,=\,E/\gamma_t \ee
with
\bd
\xi={p_1-p_2 \over\gamma p},\qquad\theta ={p_{x1}-p_{x2}\over p}
,\qquad\zeta={p_{z1}-p_{z2}\over p},\qquad4\chi^2=\theta^2+
\zeta^2 \ed
\par
We now assume that the momenta are almost perpendicular 
to the $\bar j$-axis in the c.\ o.\ m.\ system, which means (Eq.(9))
\bd
\xi^2(1+\gamma^2\chi^2)\ll4\chi^2\qquad \mbox{or}\qquad
\xi^2\ll(4\hmi\gamma^2\xi^2)\chi^2\approx4\chi^2
\ed
which can be written as
\bd 
(p_1\hmi p_2)^2\ll\gamma^2\Bigl((p_{x1}\hmi p_{x2})^2+(p_{z1}\hmi
p_{z2})^2\Bigr) \ed
This assumption gives
\be
|\vec p_1| \approx |\vec p_2| \approx p \ee
and one obtains
\be
\vec {\bar p}_{1,2} \;
= \; p \left(\begin{array}{c} 0 \\ 0 \\ \pm\chi \end{array}
\right)_{\bar j,\bar k,\bar l} \ee
The relative velocity $\,\beta_t$ of the c.\ o.\ m.\ system simplifies to
\be
\beta_t\,=\,\beta\,\cos\chi \ee
where $\beta$ is given by the mean value $p$ of all momenta and 
$\gamma_t$ is
\be
\gamma_t^2\,=\,{1\over 1-\beta_t^2}\,=\,{\gamma^2\over1+\beta^2
\gamma^2\chi^2} \ee
\par
After the collision the absolute values of the two momenta are
not changed in the
c.~o.~m.~system and the new directions of
the momenta can be described with help of the two angles
$\bar\psi$ and $\bar\phi$, where $\bar\psi$ is the angle between 
the momentum of the scattered particle $\vec{\bar p}_1^{\,\prime}$
and the (longitudinal) $\bar j$-axis, and $\bar\phi$ is the angle 
between the projection of $\vec {\bar p}_1^{\,\prime}$ on the
$\bar\ell$-$\bar k$-plane and the $\bar k$-axis. The momenta after
the collision are then given by
\be
\vec{\bar p}_{1,2}^{\,\prime}\;=\pm\bar p\left(\begin{array}{c}\cos\bar\psi
\\ \sin\bar\psi\cos\bar\phi \\
\sin\bar\psi\sin\bar\phi \end{array}\right)_{\bar j,\bar k,\bar l}\,
=\pm p\sin\chi\left(\begin{array}{c}\cos\bar\psi \\
\sin\bar\psi\cos\bar\phi \\ \sin\bar\psi\sin\bar\phi \end{array}
\right)_{\bar j,\bar k,\bar l} \ee
The inverse Lorentz transformation gives the two rotated momenta
in the laboratory system:
\be
\vec p_{1,2}^{\,\prime} 
= \left(\begin{array}{c} \gamma_t(\pm\bar p'_j+\beta_t\bar E/c) \\
\pm\bar p'_k \\ \pm\bar p'_l \end{array}\right) _{j,k,l}
=\,p\,\left(\begin{array}{c} \pm\gamma_t\sin\chi\cos\bar\psi +\cos\chi \\
\pm\sin\chi\sin\bar\psi\,\cos\bar\phi \\
\pm\sin\chi\sin\bar\psi\sin\bar\phi \end{array}\right)_{j,k,l}
\ee
The change of the momenta due to the collision is
\be
\vec p_{1,2}^{\,\prime} - \vec p_{1,2}\,=\,\pm p\sin\chi\left(
\begin{array}{c} \gamma_t\cos\bar\psi \\
\sin\bar\psi\,\cos\bar\phi \\
\sin\bar\psi\sin\bar\phi-1 \end{array}\right) _{j,k,l}
\approx\, \pm p\,\chi\left(\begin{array}{c} \gamma_t\cos\bar\psi \\
\sin\bar\psi\,\cos\bar\phi \\
\sin\bar\psi\sin\bar\phi-1 \end{array}\right)_{j,k,l} \ee
With $p_{1,2}^{\,\prime}=\sqrt{p_{j1,2}^{\,\prime2}+p_{k1,2}^{\,\prime2}+
p_{l1,2}^{\,\prime2}}\approx p_{j1,.2}^{\,\prime}+(p_{k1,2}^{\,\prime2}+
p_{l1,2}^{\,\prime2})/(2p_{j1,2}^{\,\prime})$ we get finally
\bea
p_{1,2}^\prime-p_{1,2} \,&\approx&\,p_{j1,2}^\prime-p_{j1,2} \nn \\
&\approx&\,\pm\,p\,\gamma_t\chi\cos\bar\psi  \eea

\vskip 30pt
\par\noindent
{\bf 3.\ \ \,Scattering cross section}
\vskip 10pt
\par
The probability for scattering of one of the two colliding particles
into the solid angle $d\bar\Omega$ is given by the M{\o}ller scattering 
cross-section in the c.\ o.\ m. system
$\lb8\rb$:
\be
d\bar\sigma\,=\,{r_p^2 \over 4{\bar\gamma}^2}\left(\Bigl(1+{1\over
\bar\beta^2}\Bigr)^2\Bigl({4\over\sv\bar\Phi}-{3\over\sz \bar
\Phi} \Bigr)+{4\over\sz\bar\Phi}+1 \right)d\bar\Omega \ee
$r_p$ is the classical particle radius, $\bar\gamma$ is
$\gamma/\gamma_t$, and $\bar\Phi$ is the angle between the momenta
before and after the collision, i.\ e.\ between the
$\bar\ell$-axis and $\vec{\bar p}_1^\prime$. In order to determine all
collisions with an energy change larger than the maximum stable energy
deviation we have to integrate over $\bar\psi$ and $\bar\phi$ with the 
conditions
\bd
0\,\leq\,\bar\phi\,\leq\,2\pi\,,\qquad\qquad 0\,\leq\,\bar\psi \,\leq \,
\bar\psi_m
\ed
where $\bar\psi_m$ is given by the maximum stable momentum deviation
$\Delta p_m$. If one of the two particles is scattered into the region 
$\pi\hmi\bar\psi_m\leq\bar\psi\leq\pi$ the other one is scattered into
$0\leq\bar\psi\leq\bar\psi_m$ so that both regions are included.
With Eq.(18) we obtain for $\Delta p_m$
\newpage
\vskip-20pt
\bd
\Delta p_m\,=\,|p_{1,2}^{\,\prime}-p_{1,2}|_m\,=\,p\chi\gamma_t\cos
\bar\psi_m \ed
or
\be
\cos\bar\psi_m\,=\,{\delta_m\over\gamma_t\chi}\,=\,
{\delta_m\sqrt{1+\beta^2\gamma^2\chi^2}\over\gamma\chi}\,\le\,1
\ee
with
\bd
\delta_m\,=\,{\Delta p_m\over p} \ed
Eq.(20) gives also the condition
\be
\chi^2\,\geq\,\chi_m^2\,=\,{\delta_m^2\over\gamma^2(1\hmi\beta^2
\delta_m^2)}\,\approx\,{\delta_m^2\over\gamma^2} \ee
With $d\bar\Omega=\sin\bar\psi\,d\bar\phi\,d\bar\psi$ one gets
for the total cross section
\be
\bar\sigma\,=\,{r_p^2\over4\bar\gamma^2}\int_0^{\bar\psi_m}
\int_0^{2\pi}\left(\Bigl(1+{1\over\bar\beta^2}\Bigr)^2\Bigl(
{4\over\sv\bar\Phi}-{3\over\sz\bar\Phi}\Bigr)+{4\over\sz\bar
\Phi}+1\right)\,\sin\bar\psi\,d\bar\phi\,d\bar\psi \ee
With the relation
\bd
{\bar p}_\ell^\prime\,=\,\bar p\cos\bar\Phi\,=\,\bar p\sin
\bar\psi\cos\bar\phi \ed
which follows from the definitions of the angles $\bar\Phi$,
$\bar\psi$, and $\bar\phi$, one obtains
\bd
\sz\bar\Phi\,=\,\sz\bar\phi+\cz\bar\psi\cz\bar\phi \ed
Substituting $\,\tan\bar\phi=|\cos\bar\psi|\times\tan u\,$
and integrating with respect to $u$ and $\bar\psi$ gives
\bea
\bar\sigma\,&\,=\,&\,{\pi r_p^2\over2\bar\gamma^2}
\int_0^{\bar\psi_m} \left(\Bigl(1+{1\over\bar\beta^2}\Bigr)^2
\Bigl({2(1+\cz \bar\psi) \over\cd\bar\psi}-{3\over\cos\bar\psi}
\Bigr)+{4\over\cos\bar\psi}+1\right)\,\sin\bar\psi\,d\bar\psi \nn \\
&\,=\,&\,{\pi r_p^2\gamma_t^2\over2\gamma^2}\left(\Bigl(3-{2\over
\bar\beta^2}-{1\over\bar\beta^4}\Bigr)\ln{\gamma_t\chi\over \delta_m}
+\Bigl(1+{1\over\bar\beta^2}\Bigr)^2{\gamma_t^2\chi^2-\delta_m^2 \over
\delta_m^2} +1-{\delta_m\over\gamma_t\chi}\right)
\eea
where $\bar\beta$ is given by
\be
\bar\beta^2\,=\,{\bar p^2c^2\over\bar E^2}\,=\, \,\gamma_t^2
\beta^2 \chi^2\,=\,{\beta^2\gamma^2\chi^2\over1+\beta^2\gamma^2
\chi^2}\,\approx\,{\beta^2\gamma^2\chi^2\over1+\gamma^2\chi^2} 
\ee
since $\chi^2\ll1$. The cross section $\bar\sigma\,$  is parallel to 
$\vec p_1+\vec p_2$, i.\ e.\ parallel to the $\bar\ell$-axis, and it 
is transformed into the laboratory system by:
\be
\sigma\,=\,{\bar\sigma\over\gamma_t}  \ee

\newpage
\par\noindent
{\bf 4.\ \ \,Loss rate}
\vskip 10pt
\par
We want to calculate the number of collisions per unit time which
lead to the loss of both particles. We shall integrate, therefore,
the cross section with respect to all positions and angles of the two
colliding particles satisfying the condition Eq.(21). The distribution 
of the positions and angles of electrons and positrons is, in good 
approximation, a Gaussian distribution:
\be
P_\beta(x_{\beta},x'_{\beta}, z_{\beta},z'_{\beta}) 
={\beta_x\beta_z\over4\pi^2\sigma_{x\beta}^2\sigma_{z\beta}^2 }\,
\exp\Bigl\lbrace
-{x_\beta^2+(\alpha_xx_\beta+\beta_xx'_\beta)^2\over2\sigma_{x\beta}^2}
-{z_\beta^2+(\alpha_zz_\beta+\beta_zz'_\beta)^2\over2\sigma_{z\beta}^2}
\Bigl\rbrace \ee
with $\alpha_{x,z}=-\beta_{x,z}^\prime/2$. $\sigma_{x\beta}$ and
$\sigma_{z\beta}$ are the standard deviations for the horizontal
and vertical betatron distribution. The distribution of the
synchrotron coordinates $\Delta s$ and $\Delta p$ is given by
\be
P_s(\Delta s,\Delta p)\,=\,
\, \frac {1}{2\pi\sigma_s p\,\sigma_p}\,\exp\Bigl\lbrace
-\frac {\Delta^2s}{2\sigma^2_s}-\frac 1{2\sigma^2_p} \frac
{\Delta^2p}{p^2}\Bigl\rbrace
%-\frac{1}{2\sigma^2_p}\frac{\Delta^2p}{p^2}\Bigl\rbrace
\ee
where $\sigma _p$ and $\sigma_s$ determine the relative momentum
spread and the bunch length, respectively.
\par
The number of scattering events per unit time for a single particle
moving with an angle of $2\chi$ with respect to the momentum of the 
opposing particle is $ \beta_{rel}cP_o\sigma $ where the relative 
velocity $\beta_{rel}c$ between the two colliding particles is 
$2\beta c\sin\chi \approx2\beta c\chi$.
$P_o$ is the spatial density in the laboratory system.
The total number of scattering events per unit time is obtained by 
integrating over all positions and angles of all particles with the 
condition Eq.(21):
\be
R\,=\,2\beta c\int P\sigma\,\chi\,dV \ee
$P$ and $dV$ are given by
\be
P\,=\,N_p^2
P_s(\Delta s_1,\Delta p_1)\,P_s(\Delta s_2,\Delta p_2)\,
P_\beta (x_{\beta 1},x'_{\beta 1},z_{\beta 1},z'_{\beta 1})\,
P_\beta (x_{\beta 2},x'_{\beta 2},z_{\beta 2},z'_{\beta 2})
\ee
and
\be
dV\;=\;d\Delta s_1\,dx_{\beta1}\,dz_{\beta1}\,d\Delta p_1\,
d\Delta p_2\,dx'_{\beta1}\,dx'_{\beta 2}\,dz'_{\beta1}\,
dz'_{\beta2} \ee
\vskip 1.5pt \noindent
$N_p$ is the number of particles per bunch. When averaging the
position coordinates over the whole beam we assume that the two
colliding particles have always the same position, i.\ e.\ we take
into account the following conditions:
\bd
\Delta s_1=\Delta s_2,\qquad\; x_{\beta 1}+D_x{\Delta
p_1\over p}= x_{\beta 2}+D_x{\Delta p_2\over p},\qquad\;
z_{\beta 1}+D_z{\Delta p_1\over p}= z_{\beta 2} +D_z{\Delta
p_2\over p} \ed
The integration in Eq.(28) is carried out in Appendix A2
and yields ($\tau=\beta^2\gamma^2\chi^2$)
\bea
R&=&{r_p^2c\beta_x\beta_z\sigma_hN_p^2\over8\sqrt{\pi}\beta^2
\gamma^4\sigma_{x\beta}^2\sigma_{z\beta}^2\sigma_s\sigma_p}
\int_{\tau_m}^\infty\biggl(\Bigl(2+{1\over\tau}\Bigr)^2
\Bigl({\tau/\tau_m\over1+\tau}-1\Bigr) +1-{\sqrt{
1+\tau}\over\sqrt{\tau/\tau_m}} \nn \\
&  & \hskip135pt -{1\over2\tau}\Bigl(4+{1\over\tau}\Bigr)
\ln{\tau/\tau_m\over1+\tau}\biggr)e^{-B_1\tau}I_o(B_2\tau)
\,{\sqrt{\tau}\,d\tau\over\sqrt{1+\tau}}\ \ \ \ \eea
$I_o$ is the modified Bessel function and the other quantities 
are given by
\bea
{1\over\sigma_h^2}\,&=&\,{1\over\sigma_p^2}+{D_x^2+\tilde D_x^2
\over\sigma_{x\beta}^2}+{D_z^2+\tilde D_z^2\over
\sigma_{z\beta}^2} \nn \\
&=&\,{1\over\sigma_p^2\sigma_{x\beta}^2\sigma_{z\beta}^2}\Bigl(
\tilde\sigma_x^2\sigma_{z\beta}^2+\tilde\sigma_z^2\sigma_{x\beta}^2
-\sigma_{x\beta}^2\sigma_{z\beta}^2\Bigr) 
\eea
\be
B_1\,=\,{\beta_x^2\over2\beta^2\gamma^2\sigma_{x\beta}^2}\Bigl(1-
{\sigma_h^2\tilde D_x^2\over\sigma_{x\beta}^2}\Bigr)+{\beta_z^2\over2
\beta^2\gamma^2\sigma_{z\beta}^2}\Bigl(1-{\sigma_h^2\tilde D_z^2\over
\sigma_{z\beta}^2}\Bigr) \ee
\bea
B_2^2\,&\,=\,&\,{1\over4\beta^4\gamma^4}\biggl({\beta_x^2\over 
\sigma_{x\beta}^2}\Bigl(1-{\sigma_h^2\tilde D_x^2\over
\sigma_{x\beta}^2}\Bigr)-{\beta_z^2\over\sigma_{z\beta}^2}\Bigl(
1-{\sigma_h^2\tilde D_z^2\over\sigma_{z\beta}^2}\Bigr)\biggr)^2+
{\sigma_h^4\beta_x^2\beta_z^2\tilde D_x^2\tilde D_z^2\over\beta^4
\gamma^4\sigma_{x\beta}^4\sigma_{z\beta}^4} \nn \\
&\,=\,&\,B_1^2-{\beta_x^2\beta_z^2\sigma_h^2\over\beta^4\gamma^4
\sigma_{x\beta}^4\sigma_{z\beta}^4\sigma_p^2}\Bigl(\sigma_x^2\sigma_z^2
-\sigma_p^4D_x^2D_z^2\Bigr)  \eea
\be
\tau_m =\beta^2\delta_m^2 \ee
In order to simplify the representation we have introduced
\be
\tilde D_{x,z}\,=\,\alpha_{x,z}D_{x,z}+\beta_{x,z}D_{x,z}^{'} \ee
and
\be
\tilde\sigma_{x,z}^2\,=\,\sigma_{x,z}^2+\sigma_p^2\tilde D_{x,z}^2\,=\,
\sigma_{x\beta,z\beta}^2+\sigma_p^2(D_{x,z}^2+\tilde D_{x,z}^2) \ee

\vskip 30pt
\par\noindent
{\bf 6.\ \ \,Touschek lifetime}
\vskip 10pt
The number of particles lost per unit time is given by
\be
{dN_p\over dt}\,=\,-R\,=\,-aN_p^2 \ee
The fact that always two particles are lost is taken into account
by averaging over both particles, 1 and 2, in Eqs.(28) to (30),
so that there are always two contributions to the scattering rate which 
differ only by the indices 1 and 2. Integration of Eq.(38) with respect 
to $t$ gives
\bd
-{1\over N_p}\,=\,-at+\mbox{const.} \ed
or
\be
N_p\,=\,{1\over at-\mbox{const.}}\,=\,{N_o\over1\hpl
N_oat} \ee
where $N_o$ is the number of particles at the time $t=0$. A
lifetime $T_{\ell}$ can be defined by
\be
{1\over T_{\ell}}\,=\,\left\langle\,aN_o\right\rangle\,=\,
\left\langle{R\over N_o}\right\rangle \ee
after that the number of particles drops to half the initial number.
The brackets denote the average over the whole circumference of the
storage ring. Using the same convention as other calculations we write 
\bea
{1\over T_{\ell}}\,
&=&\,\left\langle{r_p^2cN_p\over\,8\pi\gamma^2\sigma_s\sqrt{
\sigma_x^2\sigma_z^2 -\sigma_p^4D_x^2D_z^2}\,\tau_m}\,F(\tau_m,B_1,B_2) 
\right\rangle \eea
with
\bea
 F(\tau_m,B_1,B_2)\,&=&\,\sqrt{\pi(B_1^2\hmi B_2^2)}\,\tau_m
\int_{\tau_m}^\infty\biggl(\Bigl(2+{1
\over\tau} \Bigr)^2 \Bigl({\tau/\tau_m\over1+\tau}-1\Bigr)
+1-{\sqrt{ 1+\tau}\over\sqrt{\tau/\tau_m}} \qquad\quad \nn \\
&\,& \hskip80pt -{1\over2\tau}\Bigl(4+{1\over\tau}\Bigr)
\ln{\tau/\tau_m\over1+\tau}\biggr)e^{-B_1\tau}I_o(B_2\tau) \,
{\sqrt{\tau}\,d\tau\over\sqrt{1+\tau}} \ \ \ \ \eea
where $B_1^2\hmi B_2^2$ is given by Eq.(34).
A faster numerical integration is achieved by substituting $
\tau=\tan^2\kappa\,,\, \tau_m=\tan^2\kappa_m $:
\bea
 F(\tau_m,B_1,B_2)\,&=&\,2\sqrt{\pi(B_1^2\hmi B_2^2)}\,\tau_m
\int_{\kappa_m}^{\pi/2}\biggl(\bigl(2\tau\hpl1\bigr)^2
\Bigl({\tau /\tau_m\over1\hpl\tau}\hmi1\Bigr)/\tau+\tau-\sqrt{\tau
\tau_m (1\hpl\tau)} \nn \\
&  & \hskip120pt -\Bigl(2\hpl{1\over2\tau}\Bigr)\ln{\tau/\tau_m\over1\hpl
\tau}\biggr)e^{-B_1\tau}I_o(B_2\tau)\,\sqrt{1\hpl\tau}\,d\kappa\ \ \nn \eea
\par
Eqs.(41) and (42) describe the most general case with respect to the 
horizontal and vertical betatron oscillation, the horizontal and vertical 
dispersion, and the derivatives of the amplitude functions and dispersions.
Special cases with some simplifications will be discussed in the following 
sections.
\par
 Figures 1 to 4 show $F(\tau_m,B_1,B_2)$ as a function of $\tau_m=
\beta^2\delta_m^2\approx\delta_m^2$ for different ratios of beam height 
to width and for different energies.
\vskip 30pt
\par\noindent
{\bf 7.\ \ \,Special cases}
\vskip 5pt
\par\noindent
{\bf 7.1\ \,Plane orbit}
\vskip 10pt
\par
In case of a plane orbit and without coupling of the horizontal dispersion
one obtains
\be
D_z=\tilde D_z=0 \ee
and $\sigma_h$ is given by (Eq.(32))
\bd
{1\over\sigma_h^2}\,=\,{1\over\sigma_p^2}+{D_x^2+\tilde D_x^2
\over\sigma_{x\beta}^2}\,=\,{{\tilde\sigma}_x^2\over\sigma_p^2
\sigma_{x\beta}^2} \ed
With Eqs.(33) and (34) and with $\sigma_{z\beta}=\sigma_z$  
$B_1$ and $B_2$ simplify to
\be
B_{1,2}\,=\,{1\over2\beta^2\gamma^2}\biggl|{\beta_x^2\sigma_x^2
\over\sigma_{x\beta}^2\tilde\sigma_x^2}\pm{\beta_z^2\over
\sigma_{z\beta}^2}\biggr| \ee
and the lifetime is given by 
\bea
{1\over T_{\ell}}\,&=&\,\left\langle{r_p^2cN_p\over\,8\pi
\gamma^2\sigma_x\sigma_z\sigma_s\tau_m}\,F(\tau_m,B_1,B_2)\right\rangle 
\eea
\vskip 20pt
\par\noindent
{\bf 7.2\ \,Flat beam}
\vskip 10pt
\par
 For flat beams we assume
\be
{\sigma_{x\beta}^2\over\beta_x^2} \gg {\sigma_{z\beta}^2\over
\beta_z^2} \ee
and $D_z=\tilde D_z=0$.
The last condition (plane orbit) is not necessary but simplifies the
calculation. Thus we obtain with Eq.(44)
\bd
B_{1,2}\,\approx\,{\beta_z^2\over2\beta^2\gamma^2\sigma_{z\beta}^2} 
\ed
and
\bd
B_1-B_2\,=\,{\beta_x^2\sigma_x^2\over\beta^2\gamma^2\sigma_{x\beta}^2
\tilde\sigma_x^2} \ed
and with Eq.(34) and $\sigma_h=\sigma_p\sigma_{x\beta}/\tilde\sigma_x$
\bd
B_1^2-B_2^2\,=\,{\beta_x^2\beta_z^2\sigma_x^2\over\beta^4\gamma^4
\sigma_{x\beta}^2\tilde\sigma_x^2\sigma_{z\beta}^2} \ed
The argument of the Bessel function at the lower integration limit
(Eq.(42)) is ($\chi_m=\delta_m/\gamma$)
\bd
\tau_mB_2\,=\,{\delta_m^2\over2\gamma^2}\biggl|{\beta_x^2\sigma_x^2
\over\sigma_{x\beta}^2\tilde\sigma_x^2}-{\beta_z^2\over\sigma_{z\beta}^2}
\biggr|\,\approx\,{\delta_m^2\beta_z^2\over2\gamma^2\sigma_{z\beta}^2} 
\ed
If this value is large as compared to one we have
$\,\gamma^2\sigma_{z\beta}^2/\beta_z^2\ll\delta_m^2\,$. This means
that the change $\gamma_t\sigma_{z\beta}/\beta_z$ of the 
longitudinal momentum due to the initial vertical momentum
$\sigma_{z\beta}/\beta_z$ is smaller than the maximum stable
momentum deviation $\delta_m$ so that the vertical momentum does 
not contribute to the loss rate. The Bessel function can then be 
written approximately as $\lb11\rb$
\be
e^{-\tau B_1}\,I_o(\tau B_2)\,\approx\,{1\over\sqrt{2\pi\tau
B_2}}\,e^{-\tau(B_1-B_2)}\,=\,{\beta\gamma\sigma_{z\beta}\over
\sqrt{\pi\tau}\beta_z}\exp\biggl\lbrace-{\tau\beta_x^2\sigma_x^2
\over\beta^2\gamma^2\sigma_{x\beta}^2\tilde\sigma_x^2}\biggr\rbrace 
\ee
The lifetime follows with 
Eqs.(31) and (47) to 
\bea
{1\over T_\ell}\,&=&\left\langle{r_p^2c\beta_xN_p\over\,8\pi\beta\gamma^3
\sigma_{x\beta}\sigma_{z\beta}\sigma_s\tilde\sigma_x}\int_{\tau_m}^\infty
\biggl(\Bigl(2+{1\over\tau}\Bigr)^2\Bigl({\tau/\tau_m\over
1+\tau}-1\Bigr)+1-{\sqrt{1+\tau}\over \sqrt{\tau/\tau_m}}\right. \nn \\
& & \hskip120pt \left.-{1\over2\tau}\Bigl(4+{1\over\tau}\Bigr)
\ln{\tau/\tau_m\over1+\tau}\biggr)\exp\biggl\lbrace{-{\tau\epsilon_m
\over\tau_m}\biggr\rbrace}\,{d\tau\over\sqrt{1+\tau}}\right\rangle\;\;\;
\eea
and $F(\tau_m,B_1,B_2)$ (Eq.(42)) is given by
\bea
 F(\tau_m,B_1,B_2)\,&=&\,
\sqrt{{B_1^2\hmi B_2^2\over2B_2}}\,\tau_m\int_{\tau_m}^\infty\biggl(
\Bigl(2+{1\over\tau}\Bigr)^2\Bigl({\tau/\tau_m\over1+\tau}-1\Bigr)
+1-{\sqrt{1+\tau}\over\sqrt{\tau/\tau_m}} \nn \\
&\,& \hskip90pt -{1\over2\tau}\Bigl(4+{1\over\tau}\Bigr)
\ln{\tau/\tau_m\over1+\tau}\biggr)\,e^{-\tau(B_1-B_2)}\,{d\tau\over
\sqrt{1+\tau}} \nn \\
&=&\,\sqrt{\epsilon_m\tau_m}\int_{\tau_m}^\infty\biggl(\Bigl(2+
{1\over\tau}\Bigr)^2\Bigl({\tau/\tau_m\over1+\tau}-1\Bigr)+1-{\sqrt{1+
\tau}\over\sqrt{\tau/\tau_m}} \nn \\
& & \hskip70pt -{1\over2\tau}\Bigl(4+{1\over\tau}\Bigr)\ln{\tau
/\tau_m\over1+\tau}\biggr)\exp\Bigl\lbrace{-{\tau\epsilon_m
\over\tau_m}\Bigr\rbrace}\,{d\tau\over\sqrt{1+\tau}}\;\; 
\eea
where $\epsilon_m$ is a generalized expression  already used in 
other investigations:
\be
\epsilon_m \,=\,{\delta_m^2\beta_x^2\sigma_x^2\over\gamma^2
\sigma_{x\beta}^2\tilde\sigma_x^2}\,=\,{\delta_m^2\beta_x^2
\sigma_x^2\over\gamma^2\sigma_{x\beta}^2(\sigma_x^2 +\sigma_p^2
\tilde D_x^2)} \ee

\vskip 20pt
\par\noindent
{\bf 7.3\ \,Round beam}
\vskip 10pt
\par
Here we assume again $D_z=\tilde D_z=0$. Then we can define the round 
beam by
\be
{\beta_x\sigma_x\over\sigma_{x\beta}\tilde\sigma_{x\beta}}\,=\,{\beta_z
\over\sigma_{z\beta}}
\ee
and $B_1$ and $B_2$ are given by
\bd
B_1\,=\,{\beta_x^2\sigma_x^2\over\beta^2\gamma^2\sigma_{x\beta}^2
\tilde\sigma_x^2}\,=\,{\epsilon_m\over\tau_m}\,,\qquad\qquad B_2\,=\,0
\ed
$F(\tau_m,B_1,B_2)$ simplifies to
\bea
 F(\tau_m,B_1,0)\,&=&\,\sqrt{\pi}B_1\tau_m
\int_{\tau_m}^\infty\biggl(\Bigl(2+{1\over\tau}\Bigr)^2\Bigl(
{\tau/\tau_m\over1+\tau}-1\Bigr)+1-{\sqrt{ 1+\tau}\over
\sqrt{\tau/\tau_m}}\qquad\quad \nn \\
&\,& \hskip80pt -{1\over2\tau}\Bigl(4+{1\over\tau}\Bigr)\ln{
\tau/\tau_m\over1+\tau}\biggr)\exp\bigl\lbrace-\tau B_1\bigr\rbrace 
\,{\sqrt{\tau}\,d\tau\over\sqrt{1+\tau}}\ \ \ \  
\eea
With help of the integral $\lb11\rb$
\bea
\int_0^\infty e^{-p\tau}{d\tau\over\sqrt{\tau(\tau\hpl a)}}
\,&=&\,\exp\lbrace{-ap/2\rbrace}K_o(ap/2)  \eea
where $K_o$ is a Bessel function, the integral in Eq.(52) can be solved 
exactly except for the terms with $\ln(\tau/(1+\tau))$. Then one gets for
small $\tau_m$:
\be
 F(\tau_m,B_1,0)\,=\,\sqrt{\pi}B_1(2+B_1)(K_1-K_o)e^{B_1/2}+0(\sqrt{\tau_m})
\ee
where $K_1$ and $K_o$ are Bessel functions with the argument $B_1/2$.
\par 
Thus for a round beam $F$ becomes constant with decreasing $\tau_m$. This is
different in the case of a flat beam where $F$ increases continuously with
decreasing $\tau_m$. The reason for the different behaviour is the different
distribution of the transverse momentum.
When the maximum stable energy deviation $\delta_m$ is decreased by
$-d\delta_m$ more particles will be lost having the smaller transverse
momentum difference $\Delta p_\perp\hmi d\Delta p_\perp$. In a 
one-dimensional distribution the number of particles which can be 
scattered additionally is proportional to $2d\Delta p_x$, but in a 
two-dimensional distribution it is proportional to $2\pi\Delta p_\perp 
d\Delta p_\perp$ with $\Delta^2p_\perp=\Delta^2p_x+ \Delta^2p_z$. 
Thus with decreasing $\Delta p_\perp$ the increase in loss rate is 
larger in a one-dimensional distribution, i.\ e.\ in a flat beam. 
\vskip 20pt
\par\noindent
{\bf 7.4\ \,Non-relativistic case}
\vskip 10pt
\par
The non-relativistic case is defined by $\bar\beta^2\,\ll\,1$, i.\ e.\
(Eq.(24))
\be
\beta^2\gamma^2\chi^2\,\ll\,1 \ee
which means also $\gamma^2\chi^2\ll1$ since $\chi^2\ll1$. 
$\bar\psi_m$ is given by
\bd
\cos\bar\psi_m\,=\,{\delta_m\over\gamma_t\chi}\,=\,{\delta_m
\over\gamma\chi} \ed
since $\gamma_t=\gamma$ (Eq.(14)). One obtains for the cross sections 
with Eq.(23)
\bea
\bar\sigma\,&=&\,{\pi r_p^2\over2\bar\beta^4}
\left({1\over\cz\bar\psi_m}-1+\ln\cos\bar\psi_m\right) \nn \\
&=&\,{\pi r_p^2\over2\beta^4\gamma^4\chi^4} \left({\gamma^2
\chi^2\over\delta^2_m}-1+\ln{\delta_m\over\gamma\chi}\right) 
\eea
and with $\chi^2=\tau/(\beta^2\gamma^2)$
\bea
\sigma\,&=&\,{\pi r_p^2\over\,2\gamma\tau^2}\left(
{\tau\over\beta^2\delta_m^2}-1+\ln{\beta\delta_m\over\sqrt{\tau}}
\right) \eea
With Eqs.(40), (55) and (A2.5) one gets for the lifetime
\bea
{1\over T_\ell}\,&=&\,\left\langle{cr_p^2\beta_x\beta_z\sigma_hN_p\over
8\sqrt{\pi}\beta^2\gamma^4\sigma_{x\beta}^2\sigma_{z\beta}^2
\sigma_s\sigma_p}\int_{\tau_m}^\infty\left({\tau\over\tau_m}
-1-{1\over2}\ln{\tau\over\tau_m}\right)\,e^{-\tau B_1}\,I_o
\bigl(\tau B_2\bigr) \,{d\tau\over\tau^{3/2}}\right\rangle 
\eea
and
\bea
 F(\tau_m,B_1,B_2)\,=\,\sqrt{\pi(B_1^2-B_2^2)}\,\tau_m
\int_{\tau_m}^\infty\left({\tau\over\tau_m}
-1-{1\over2}\ln{\tau\over\tau_m}\right)\,e^{-\tau B_1}\,I_o
\bigl(\tau B_2\bigr) \,{d\tau\over\tau^{3/2}} \eea
One can obtain this result also from Eq.(42) by neglecting $\tau$
as compared one.
Using the same notation as $\lb9\rb$ one can write with 
$D_z=\tilde D_z=0$ 
\bea
{1\over T_\ell}\,&=&\,\left\langle{cr_p^2\beta_x^2\beta_z\sigma_xN_p
\over8\sqrt{\pi} \beta^3\gamma^5\tilde\sigma_x^2\sigma_{x\beta}^2
\sigma_{z\beta}^2\sigma_s}\int_{\epsilon_m}^\infty\left({
\epsilon\over\epsilon_m}\hmi1\hpl{1\over2}\ln{\epsilon_m
\over\epsilon}\right)e^{-\epsilon G_1}I_o(\epsilon G_2){d\epsilon
\over\epsilon^{3/2}}\right\rangle \eea
with
\bd
G_{1,2}\,=\,{\beta^2\gamma^2\sigma_{x\beta}^2
\sigma_{z\beta}^2\over\beta_x^2\beta_z^2}\,B_{1,2}\,
=\,{1\over2}\biggl|{\sigma_{z\beta}^2\sigma_x^2\over
\beta_z^2\tilde\sigma_x^2}\pm{\sigma_{x\beta}^2
\over\beta_x^2}\biggr| \ed

\vskip 20pt
\par\noindent
{\bf 7.5\ \,Non-relativistic case for a flat beam}
\vskip 10pt
 For the non-relativistic case one obtains with Eqs.(47) and (58) 
and with $\bar\beta^2  = \tau$, $\gamma_t^2
\chi^2 = \tau/\beta^2$, $\chi=\sqrt{\tau}/
(\beta\gamma)$, and $D_z=\tilde D_z=0$
\bea
{1\over T_\ell}\,&=&\,\left\langle{r_p^2c\beta_xN_p\over8\pi\beta
\gamma^3\tilde\sigma_x\sigma_{x\beta}\sigma_{z\beta}\sigma_s}
\int_{\tau_m}^\infty\biggl({\tau\over\tau_m}-1+{1\over2}
\ln{\tau_m\over\tau}\biggr)\exp\Bigl\lbrace{-{\tau\epsilon_m
\over\tau_m}\Bigr\rbrace}\,{d\tau\over\tau^2}\right\rangle 
\eea
and
\be
 F(\tau_m,B_1,B_2)\,=\,{\beta\gamma\sigma_{z\beta}\tau_m\over
\sqrt{\pi}\beta_z}\int_{\tau_m}^\infty\biggl({\tau\over\tau_m}-1-
{1\over2}\ln{\tau\over\tau_m}\biggr)\exp\Bigl\lbrace{-{\tau
\epsilon_m\over\tau_m}\Bigr\rbrace}\,{d\tau \over\tau^2}
\ee
 Following existing representations (see f.\ e.\ $\lb9,10\rb$) we can 
write $\bigl(\tau=\epsilon\tau_m/\epsilon_m$, $\epsilon_m/\tau_m=\beta_x^2
\sigma_x^2/(\beta^2\gamma^2\sigma_{x\beta}^2\tilde\sigma_x^2)\bigr)$
\bea
{1\over T_\ell}\,&=&\,\left\langle{r_p^2c\beta_x^3\sigma_x^2N_p\over
8\pi\beta^3\gamma^5\sigma_{x\beta}^3\sigma_{z\beta}\sigma_s
{\tilde\sigma_x}^3}\int_{\epsilon_m}^\infty\biggl({\epsilon
\over\epsilon_m}-1-{1\over2}\ln{\epsilon\over\epsilon_m}\biggr)\exp
\lbrace{-\epsilon\rbrace}\,{d\epsilon\over\epsilon^2}\right\rangle \nn \\
&=&\,\left\langle{r_p^2c\beta_x^3\sigma_x^2N_p\over8\pi\beta^3
\gamma^5\sigma_{x\beta}^3\sigma_{z\beta}\sigma_s{\tilde\sigma_x}^3
\epsilon_m}\,C(\epsilon_m)\right\rangle \nn \\
&=&\,\left\langle{r_p^2c\beta_xN_p\over8\pi\beta^3\gamma^3\sigma_{x\beta}
\sigma_{z\beta}\sigma_s{\tilde\sigma_x}\delta_m^2}\,C(\epsilon_m)
\right\rangle 
\eea
with
\bea
C(\epsilon_m)\,&\,=\,&\,\epsilon_m\,\int_{\epsilon_m}^\infty
\biggl({\epsilon\over\epsilon_m}-1-{1\over2}\ln{\epsilon\over
\epsilon_m}\biggr)\,e^{-\epsilon}\,{d\epsilon\over\epsilon^2} \nn \\
&=&\,-{3\over2}\,e^{-\epsilon_m}+\int_{\epsilon_m}^\infty
\biggl(1+{3\epsilon_m\over2}+{\epsilon_m\over2}\ln{\epsilon\over
\epsilon_m}\biggr)\,e^{-\epsilon}\,{d\epsilon\over\epsilon}\eea
or
\bd
C(\epsilon_m)\,=\,{\sqrt{\pi}\beta_z\over\beta\gamma
\sigma_{z\beta}}\,F(\tau_m,B_1,B_2) \ed
 For $\tilde D_x=0$ and $\tilde\sigma_x=\sigma_x$ this representation
was obtained in $\lb9\rb$.

\vskip 20pt
\par\noindent
{\bf 7.6\ \,Ultra-relativistic case}
\vskip 10pt
The ultra-relativistic case is defined by
\bd
\beta^2\gamma^2\chi^2\,\gg\,1 \ed
which means $\beta^2\approx1$ since $\chi^2\ll1$. With Eqs.(14) and (20)
one obtains
\bd
\gamma_t\,=\,{1\over\chi}\,,\qquad\quad\cos\bar\psi_m\,=\,
\delta_m \ed
The cross section follows from Eq.(23) with $\bar\beta^2=1-1/\bar\gamma^2
=1-1/(\gamma^2\chi^2)\approx1$ to
\bea
\bar\sigma\,
&=&\,{\pi r_p^2\over2\gamma^2\chi^2}\left(4\,{1-\delta_m^2
\over\delta_m^2}+1-\delta_m+{4\over\gamma^2\chi^2}\ln\delta_m
\right) \nn \\
&\approx&\,{2\pi r_p^2\over\gamma^2\chi^2\delta_m^2}
\eea
With Eqs.(25), (40) and (A2.5) and with $\beta$\,=\,1 one obtains
\bea
{1\over T_\ell}\,&=&\,\left\langle{cr_p^2\beta_x\beta_z\sigma_hN_p\over
2\sqrt{\pi}\gamma^4 \sigma_{x\beta}^2\sigma_{z\beta}^2\sigma_s
\sigma_p\delta_m^2} \int_{\delta_m^2}^\infty
\exp\lbrace{-\tau B_1\rbrace}\,I_o \bigl(\tau B_2\bigr)\,d\tau
\right\rangle 
\eea
Using a definite integral $\lb11\rb$ we can write
\bea
{1\over T_\ell}\,&=&\,\left\langle{cr_p^2\beta_x\beta_z\sigma_hN_p\over
2\sqrt{\pi}\gamma^4\sigma_{x\beta}^2\sigma_{z\beta}^2\sigma_s\sigma_p
\delta_m^2}\biggl({1\over\sqrt{B_1^2\hmi B_2^2}}-\int_{0}^{\delta_m^2}
\exp\lbrace{-\tau B_1\rbrace}\,I_o \bigl(\tau B_2\bigr)\,
d\tau\biggr)\right\rangle \nn \\
&\approx&\,\left\langle{cr_p^2N_p\over2\sqrt{\pi}\gamma^2\sigma_s
\sqrt{\sigma_x^2\sigma_z^2-\sigma_p^4D_x^2D_z^2}\,\delta_m^2}\right\rangle
\ \ \eea
and $F(\tau_m,B_1,B_2)$ simplifies to
\be
 F(\tau_m,B_1,B_2)\,=\,4\sqrt{\pi} \ee

\vskip 20pt
\par\noindent
{\bf 7.7\ \,Ultra-relativistic case for a flat beam}
\vskip 10pt
A better approximation for a flat beam at high energy can be derived
by using Eq.(49). Since $\beta_{x}^2/(\gamma^2\sigma_{x\beta}^2)$ $\ll$ 
1 and $\beta_{z}^2/(\gamma^2\sigma_{z\beta}^2)$ $\gg$ 1 it follows that
$B_{1,2} \gg 1$ but $B_1\hmi B_2 \ll 1$. Eq.(49) can then be written  
approximately as
\bea
 F(\tau_m,B_1,B_2)\,&=&\,
\sqrt{B_1\hmi B_2}\,\int_{\tau_m}^\infty\biggl(
\Bigl(4\hmi3\,\tau_m\hmi\tau^{3/2}\sqrt{1\hpl
1/\tau}\,\Bigr)\,e^{-\tau(B_1-B_2)} \qquad\quad \nn \\
&\,&\hskip30pt -{4\over1\hpl\tau}+\Bigl({4\over\tau}
\hpl{1\over\tau^2}\Bigr)\Bigl({\tau\over1\hpl\tau}\hmi\tau_m\hmi{\tau_m\over2}
\ln{\tau/\tau_m\over1\hpl\tau}\Bigr)\biggr)\,{d\tau\over\sqrt{1\hpl\tau}} 
\qquad\eea
 For small $\tau_m$ and small $B_1\hmi B_2$ the function $F$ 
simplifies to

\be
 F(\tau_m,B_1,B_2)\,=\,\sqrt{\pi}\,\Bigl(4+2\,(B_1\hmi B_2)
\Bigr)+\sqrt{B_1\hmi B_2}\,\Bigl(\,\ln{(4/\tau_m)}\hmi11\Bigr) \ee

\par\noindent
where $B_1\hmi B_2$ is given by $\beta_x^2\sigma_x^2/(\gamma^2
\sigma_{x\beta}^2\tilde\sigma_x^2)$.
 For $\tau_m\!<\!10^{-3}\,(\delta\!<\!3.2\%)$ and $B_1\hmi B_2\!<\!0.3$
$(\gamma\!>\!2\beta_x/\sigma_{x\beta})$ the error is smaller than 
$8*10^{-3}$. 

Eq.(70) shows that for decreasing $\tau_m$ $F$ increases continuously, and 
that the gradient of the increase goes to zero with increasing energy.
\newpage
\par\noindent
{\bf 7.8\ \,Ultra-relativistic case for a round beam}
\vskip 10pt
In this case both,  
 $\beta_{x}^2/(\gamma^2\sigma_{x\beta}^2)$ and
 $\beta_{z}^2/(\gamma^2\sigma_{z\beta}^2)$, and therefore also $B_1$
are small as compared to one ($B_2=0$). With approximations
for the Bessel functions $K_o$ and $K_1$ $\lb11\rb$ and with Eqs.(52) 
and (54) one obtains
\bea
 F(\tau_m,B_1, 0)\,&=&\sqrt{\pi}\,\Bigl(4+B_1\bigl(\,1.73+2\,\ln(B_1)-8
\sqrt{\tau_m}\,\bigr)\Bigr)
\eea
where $B_1$ is given by $\beta_x^2\sigma_x^2/(\gamma^2\sigma_{x\beta}^2
\tilde\sigma_x^2)$. 
 For $\tau_m<10^{-3}\,(\delta<3.2\%)$ and $B_1 < 0.1\;(\gamma\!>\!3
\beta_x/\sigma_{x\beta})$ the error is smaller than $3*10^{-3}$. 
\vskip 20pt
\centerline{\bf *\quad\,\,*\quad\,\,*}
\vglue 5.0cm

\newpage
\par\noindent
{\bf Appendix A1}
\vskip 10pt
\par
Since in the c.\ o.\ m.\ system $\bar E_1=\bar E_2$ Eq.(6) can be
written as
\bea
\bar E_{1,2}&=&(\bar E_1+\bar E_2)/2 \nn \\
&=&\gamma_t\bigl(E_1\hpl E_2\hmi\beta_t(\beta_1 E_1
\cos\chi_1\hpl\beta_2E_2\cos\chi_2)\bigr)/2 \nn \\
&=&\gamma_t\bigl(E_1\hpl E_2\hmi\beta_t^2(E_1\hpl E_2)\bigr)/2 \nn \\
&=&(E_1\hpl E_2)/(2\gamma_t) \nn \\
&\approx&\,E/\gamma_t \nn  \eea  
\vskip-26pt
\hfill (A1.1)
\vskip8pt
\par\noindent
With Eq.(7) one gets
\bea
{1\over\gamma_t^2}\,&=&\,1-{c^2(\vec p_1+\vec p_2)^2\over(E_1+E_2)^2} 
\nn \\
&=&\,{E_1^2-c^2p_1^2+E_2^2-c^2p_2^2+2E_1E_2-2c^2p_1p_2\cos(\chi_1\hpl
\chi_2)\over(E_1+E_2)^2} \nn \\
&=&\,2\,{1+\gamma_1\gamma_2-\beta_1\beta_2\gamma_1\gamma_2\cos(\chi_1\hpl
\chi_2)\over(\gamma_1+\gamma_2)^2} \nn \\
&\approx&\,{1+\gamma^2-\beta^2\gamma^2(1-2\chi^2)\over2\gamma^2} \nn \\
&=&\,{1+\beta^2\gamma^2\chi^2\over\gamma^2}  \nn \eea
\vskip-30pt
\hfill (A1.2)
\vskip10pt
\par\noindent
 From the Lorentz transformation
\bd
p_{j1,2}\,=\,\gamma_t(\bar p_{\bar j1,2}+\beta_t\bar E_{1,2}/c)
\ed
 follows with Eqs.(7) and (A1.1)
\bea
\bar p_{\bar j1,2}&=&p_{j1,2}/\gamma_t-\beta_t\bar E_{1,2}/c \nn \\
&=&p_{j1,2}/\gamma_t-\beta_t(E_1\hpl E_2)/(2c\gamma_t) \nn \\
&=&(2p_{j1,2}-p_{j1}-p_{j2})/(2\gamma_t) \nn \\
&=&\pm(p_{j1}-p_{j2})/(2\gamma_t) \nn \\
&=&\pm{p_{j1}^2-p_{j2}^2\over2\gamma_t(p_{j1}+p_{j2})} \nn 
\eea
With $p_{\ell1}=p_{\ell2}$ (Eq.(4)) and with Eq.(A1.2) one gets
\bea
\bar p_{\bar j1,2}&=&\,\pm{p_1^2-p_2^2\over2\gamma_t(p_{j1}+p_{j2})} \nn \\
&=&\,\pm{\gamma\xi p(p_1+p_2)\over2\gamma_t(p_{j1}+p_{j2})} \nn \\ 
&=&\,\pm{\xi p(p_1+p_2)\over2(p_{j1}+p_{j2})}\sqrt{1\hpl\beta^2\gamma^2
\chi^2} \nn \\
&\approx&\,\pm\xi \,p\sqrt{1\hpl\beta^2\gamma^2\chi^2}/2  \nn \eea
\vskip-25pt
\hfill (A1.3)
\newpage
\par\noindent
{\bf Appendix A2}
\vskip 10pt
\par
We replace the variables $x_{\beta1,2}$, $x^\prime_{\beta1,2}$,
$z_{\beta1,2}$, $z^\prime_{\beta1,2}$, and $\Delta p_{1,2}$ by
the variables $x_\beta$, $x^\prime_\beta$, $z_\beta$,
$z^\prime_\beta$, $\Delta p$, $\xi$, $\theta$, and $\zeta$ with
help of the relations
\vskip-12pt
\bd
x_{\beta1,2}=x_\beta\mp D_x\gamma\xi/2,\qquad\qquad 
z_{\beta1,2}=z_\beta\mp D_z\gamma\xi/2 \ed
\vskip-19pt
\bd
x^\prime_{\beta1,2}=x^\prime_\beta\pm\theta/2\mp D_x^\prime
\gamma\xi/2,\quad\quad
z^\prime_{\beta1,2}=z^\prime_\beta\pm\zeta/2\mp D_z^\prime\gamma
\xi/2,\quad\quad
\Delta p_{1,2}=\Delta p\pm p \gamma\xi/2 \ed
and obtain
\vskip-15pt
\bea
R\,&=&\,{2c\beta\beta_x^2\beta_z^2N_p^2\over64\pi^6
\sigma_{x \beta}^4\sigma_{z \beta}^4\sigma_s^2\sigma_p^2p^2}\int
{\chi\sigma(\chi)}\,e^{-H}\,dV \nn \\
&=&\,{c\beta\gamma\beta_x^2\beta_z^2N_p^2\over32\pi^6
\sigma_{x \beta}^4\sigma_{z \beta}^4\sigma_s^2\sigma_p^2p}\int
{\chi\sigma(\chi)}\,e^{-H}\,dV^* \nn  \eea
\vskip-33pt
\hfill (A2.1)
\vskip18pt
\par\noindent
with the condition $\chi^2\ge\chi^2_m=\delta^2_m/\gamma^2$ (Eq.(21)) 
and with
\bea
H\,&=&\,-{x_\beta^2+(\alpha_xx_\beta+\beta_xx'_\beta)^2\over
\sigma_{x\beta}^2}-{z_\beta^2+(\alpha_zz_\beta+\beta_zz'_\beta)^2
\over\sigma_{z\beta}^2}-{1\over\sigma^2_p}{\Delta^2p\over p^2}-{\Delta^2s
\over\sigma^2_s} \nn \\
& &\hskip50pt -{\gamma^2\xi^2\over4 \sigma^2_p}
-{D_x^2\gamma^2\xi^2+(\tilde D_x\gamma\xi-\beta_x \theta)^2
\over4\sigma_{x\beta}^2}-{D_z^2\gamma^2\xi^2+(\tilde D_z \gamma
\xi-\beta_z\zeta)^2\over4\sigma_{z\beta}^2} \nn \eea
and
\bd
dV^*\,=\,d\Delta s_1\,dx_\beta\,dz_\beta\,d\Delta p\,dx_\beta^\prime
\,dz_\beta^\prime\,d\xi\,d\theta\,d\zeta
\ed
The Jacobian of the transformation is $\gamma p$. In Eq.(A2.1) six
of the nine integrations can be performed immediately and one
obtains
\bea
R\,
&=&\,{c\beta\gamma\beta_x\beta_zN_p^2\over32\pi^3\sigma_{x\beta}^2
\sigma_{z \beta}^2\sigma_s\sigma_p}\int_{-\infty}^\infty
\int_{-\infty}^\infty\int_{-\infty}^\infty\exp\Bigl\lbrace
-{\bigl(\gamma\xi-(\beta_x\tilde D_x\theta/
\sigma_{x\beta}^2+\beta_z\tilde D\zeta/\sigma_{z\beta}^2)
\sigma_h^2\bigr)^2 \over 4\sigma_h^2} \nn \\
& &\hskip65pt +{\sigma_h^2\over4}\Bigl({\beta_x\tilde D_x\theta\over
\sigma_{x\beta}^2}+{\beta_z\tilde D_z\zeta\over\sigma_{z\beta}^2}
\Bigr)^2-{\beta_x^2\theta^2\over4\sigma_{x\beta}^2}-{\beta_z^2
\zeta^2\over4\sigma_{z\beta}^2}\Bigl\rbrace {\chi\sigma(\chi)
}\,d\xi\,d\theta\,d\zeta \qquad\qquad \nn \eea
\vskip-34pt
\hfill (A2.2)
\vskip12pt
\par\noindent
with the condition $\chi^2\ge\chi^2_m$. Integration over $\xi$ gives
\bea
R\,&=&\,{c\beta\beta_x\beta_z\sigma_hN_p^2\over16\pi^{5/2} \sigma_
{x\beta}^2\sigma_{z\beta}^2\sigma_s\sigma_p}
\int_{-\infty}^\infty\int_{-\infty}^\infty{\chi\sigma(\chi)} \nn \\
& &\hskip80pt\times\exp\Bigl\lbrace {\sigma_h^2\over4}\Bigl({\beta_x
\tilde D_x\theta\over\sigma_{x\beta}^2}+{\beta_z\tilde D_z
\zeta\over\sigma_{z\beta}^2}\Bigr)^2-{\beta_x^2\theta^2\over4
\sigma_{x\beta}^2}-{\beta_z^2\zeta^2 \over4\sigma_{z\beta}^2}
\Bigl\rbrace\,d\theta\,d\zeta \qquad\qquad \nn  \eea
\vskip-34pt
\hfill (A2.3)
\vskip14pt\noindent
with the condition $ \theta^2+\zeta^2=4\chi^2\ge4\chi_m^2 $ for
the double integral. $\sigma_h$ is given by Eq.(32).  Substituting
\bd
\theta=\sqrt{\rho}\cos\nu,\qquad\zeta=\sqrt{\rho}\sin\nu,\qquad
d\theta\,d\zeta=d\rho\,d\nu/2 \ed
one obtains ($\chi=\sqrt{\rho}/2$)
\newpage
\vskip-25pt
\bea
\hskip-5pt R\!\!\!&=&\!\!\!{c\beta\beta_x\beta_z\sigma_hN_p^2\over32\pi^{5/2}
\sigma_{x\beta}^2\sigma_{z\beta}^2\sigma_s\sigma_p}
\int_{4\chi_m^2}^\infty\int_0^{2\pi}\hskip-3pt{\chi\,\sigma(\chi)}
\exp\Bigl\lbrace-{\rho\over4}\Bigl({\beta_x^2
\cz\nu\over\sigma_{x\beta}^2}\Bigl(1-{\sigma_h^2\tilde D_x^2\over
\sigma_{x\beta}^2}\Bigr) \nn \\
& &\hskip80pt +{\beta_z^2\sz\nu\over\sigma_{z\beta}^2}
\Bigl(1-{\sigma_h^2\tilde D_z^2\over\sigma_{z\beta}^2}\Bigr)
-{2\sigma_h^2\beta_x\beta_z\tilde D_x\tilde D_z\sin\nu
\cos\nu\over \sigma_{x\beta}^2\sigma_{z\beta}^2}\Bigr)
\Bigr\rbrace\,d\nu\,d\rho \nn \\
&=&\!\!\!{c\beta\beta_x\beta_z\sigma_hN_p^2\over32\pi^{5/2}
\sigma_{x \beta}^2\sigma_{z\beta}^2\sigma_s\sigma_p}
\int_{4\chi_m^2}^\infty\int_0^{2\pi}\hskip-6pt\chi{\sigma(\chi)}\exp
\bigl\lbrace\hmi\rho(A_1\hpl A_3\cos(2\nu)\hpl A_4\sin(2\nu))
\bigr\rbrace d\nu d\rho \qquad\qquad\nn   \eea
\vskip-34.5pt
\hfill (A2.4)
\vskip18pt
\par\noindent
with
\vskip-9pt
\bd
A_{1,3}\,=\,{\beta_x^2\over8\sigma_{x\beta}^2}\Bigl(1-{\sigma_h^2
\tilde D_x^2\over\sigma_{x\beta}^2}\Bigr)\pm{\beta_z^2\over8
\sigma_{z\beta}^2}\Bigl(1-{\sigma_h^2\tilde D_z^2\over
\sigma_{z\beta}^2}\Bigr)\,, \qquad
A_4\,=\,{\sigma_h^2\beta_x\beta_z\tilde D_x\tilde D_z\over
4\sigma_{x\beta}^2\sigma_{z\beta}^2} \ed
With $\cos\phi_o=A_3/\sqrt{A_3^2+A_4^2}$ and
$\sin\phi_o=A_4/\sqrt{A_3^2+A_4^2}$ and using $\lb11\rb$
\bd
I_o(\rho A_2)\,=\,{1\over\pi}\int_0^\pi e^{\pm\rho A_2\cos\theta}
\,d\theta\,=\,{1\over2\pi}\int_0^{2\pi}e^{\pm\rho A_2\cos(2\theta)}
d\theta \ed
where $I_o$ is the modified Bessel function, the double integral 
simplifies to
\bea
R\,&=&\,{c\beta\beta_x\beta_z\sigma_hN_p^2\over32
\pi^{5/2}\sigma_{x\beta}^2\sigma_{z\beta}^2\sigma_s\sigma_p}
\int_{4\chi_m^2}^\infty\int_0^{2\pi}\chi{\sigma(\chi)}
\exp\lbrace{-\rho A_1-\rho A_2\cos(2\nu-\phi_o) \rbrace}\,
d\nu\,d\rho \nn \\
&=&\,{c\beta\beta_x\beta_z\sigma_hN_p^2\over16\pi^{3/2}
\sigma_{x\beta}^2\sigma_{z\beta}^2\sigma_s\sigma_p}
\int_{4\chi_m^2}^\infty\chi{\sigma(\chi)}\exp\lbrace{-\rho
A_1\rbrace}\,I_o\bigl(\rho A_2\bigr)\,d\rho \nn \eea
\vskip-34pt
\hfill (A2.5)
\vskip13pt
\par\noindent
with
\vskip-15pt
\bea
A_2^2\,&=&\,A_3^2+A_4^2\,=\,{1\over64}\biggl({\beta_x^2\over
\sigma_{x\beta}^2}\Bigl(1-{\sigma_h^2\tilde D_x^2\over\sigma
_{x\beta}^2}\Bigr)-{\beta_z^2\over\sigma_{z\beta}^2}\Bigl(1-
{\sigma_h^2\tilde D_z^2\over\sigma_{z\beta}^2}\Bigr)\biggr)^2
+{\sigma_h^4\beta_x^2\beta_z^2\tilde D_x^2\tilde D_z^2\over
16\sigma_{x\beta}^4\sigma_{z\beta}^4} \nn \\
&=&\,A_1^2-{\beta_x^2\beta_z^2\sigma_h^2\over16
\sigma_{x\beta}^2\sigma_{z\beta}^2}\Bigl({1\over\sigma_p^2}
+{D_x^2\over\sigma_{x\beta}^2}+{D_z^2\over\sigma_{z\beta}^2} 
\Bigr) \nn \\ 
&=&\,A_1^2-{\beta_x^2\beta_z^2\sigma_h^2\over16
\sigma_{x\beta}^4\sigma_{z\beta}^4\sigma_p^2}\Bigl(
\sigma_x^2\sigma_z^2-\sigma_p^4D_x^2D_z^2 \Bigr) \nn  \eea
\vskip-33pt
\hfill (A2.6)
\vskip17pt
\par\noindent
With Eqs.(23) and (25) one gets ($\rho=4\chi^2$)
\bea
R\!&=&\!{r_p^2c\beta\beta_x\beta_z\sigma_hN_p^2\over32
\sqrt{\pi}\gamma^2\sigma_{x\beta}^2\sigma_{z\beta}^2\sigma_s
\sigma_p}\int_{4\delta_m^2/\gamma^2}^\infty\biggl(
\Bigl(1+{1\over \bar\beta^2}\Bigr)^2\Bigl({\gamma_t^2
\chi^2\over\delta_m^2}-1\Bigr)+1-{\delta_m\over\gamma_t\chi} \nn \\
& & \hskip80pt +\Bigl(3-{2\over\bar\beta^2}-{1\over\bar\beta^4}\Bigr)
\ln{\gamma_t\chi\over\delta_m}\,\biggr)\,\exp\lbrace{-\rho A_1\rbrace}
\,I_o\bigl(\rho A_2\bigr)\,\gamma_t\chi\,d\rho \qquad\quad \nn \eea
\vskip-31pt
\hfill (A2.7)
\vskip18pt\noindent 
With $\rho=4\tau/(\beta^2\gamma^2)$,
$B_{1,2}=4A_{1,2}/(\beta^2\gamma^2)$, and $\tau_m =\beta^2\delta_m^2$
one obtains finally Eq.(31).

\newpage
\begin{figure}[p]
\psfig{figure=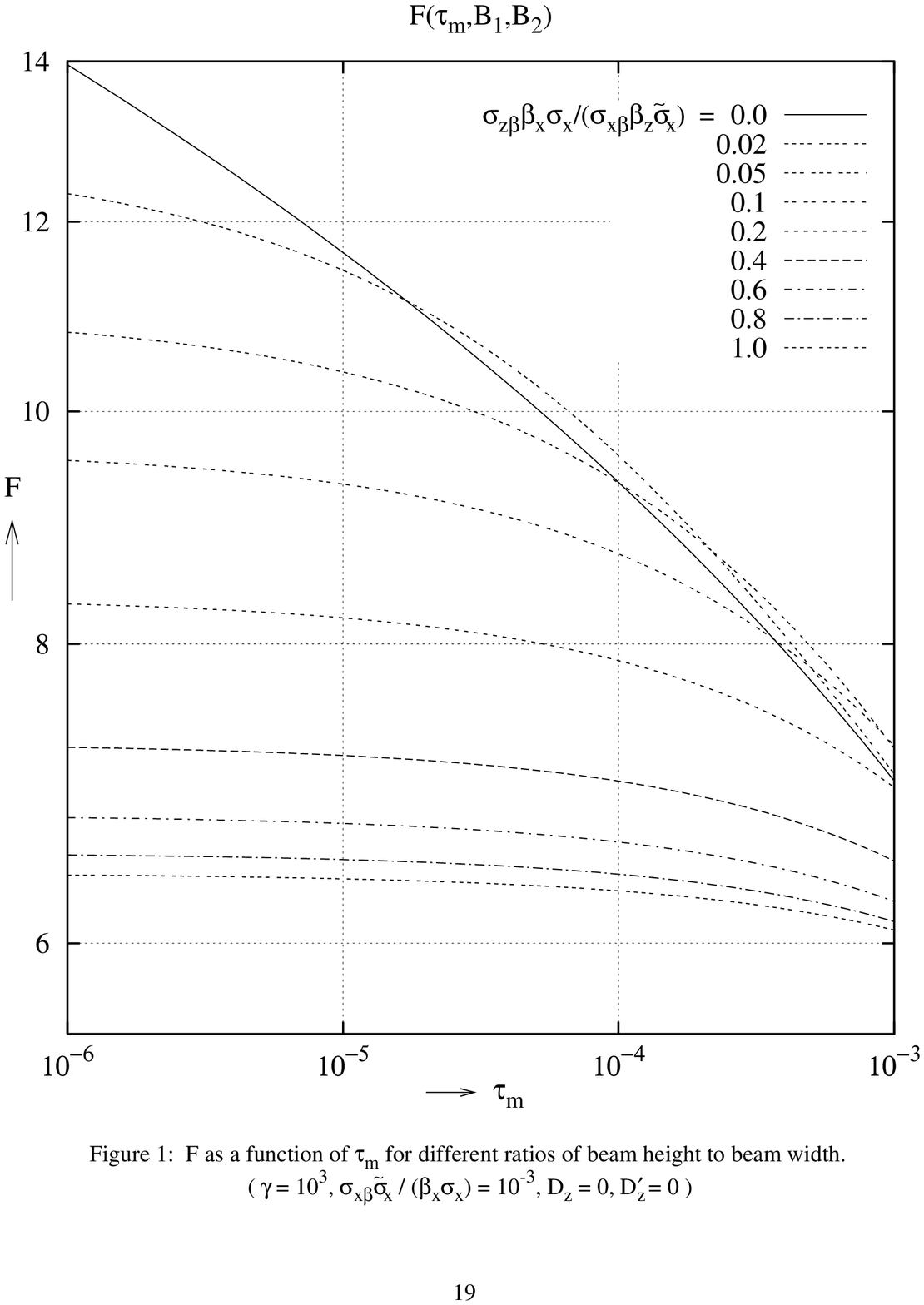,height=7.5in,width=7.0in}
\end{figure}

\newpage

\begin{figure}[p]
\psfig{figure=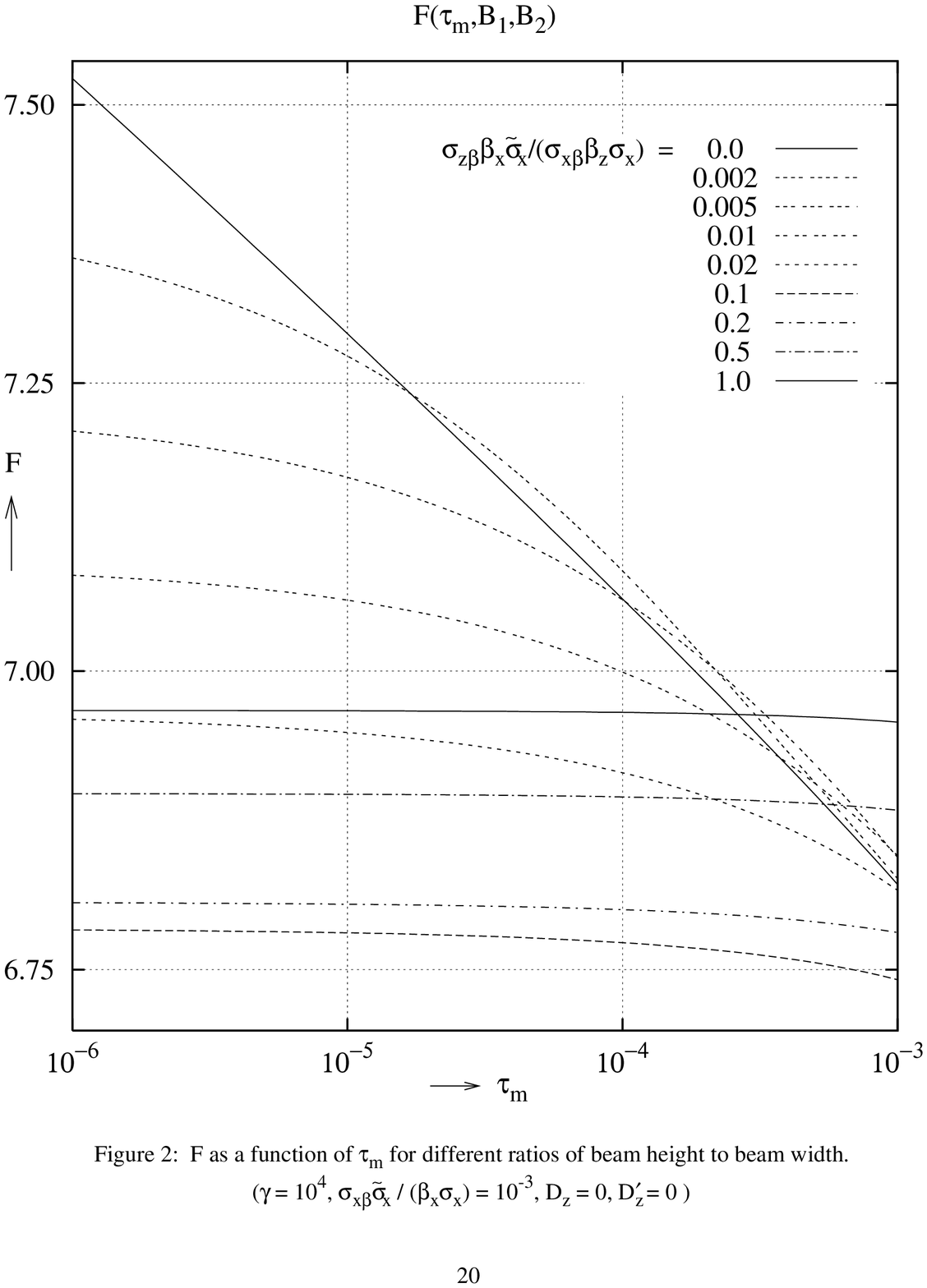,height=7.5in,width=7.0in}
\end{figure}

\newpage

\hoffset -2.truecm
\voffset -2.truecm
\begin{figure}[p]
\psfig{figure=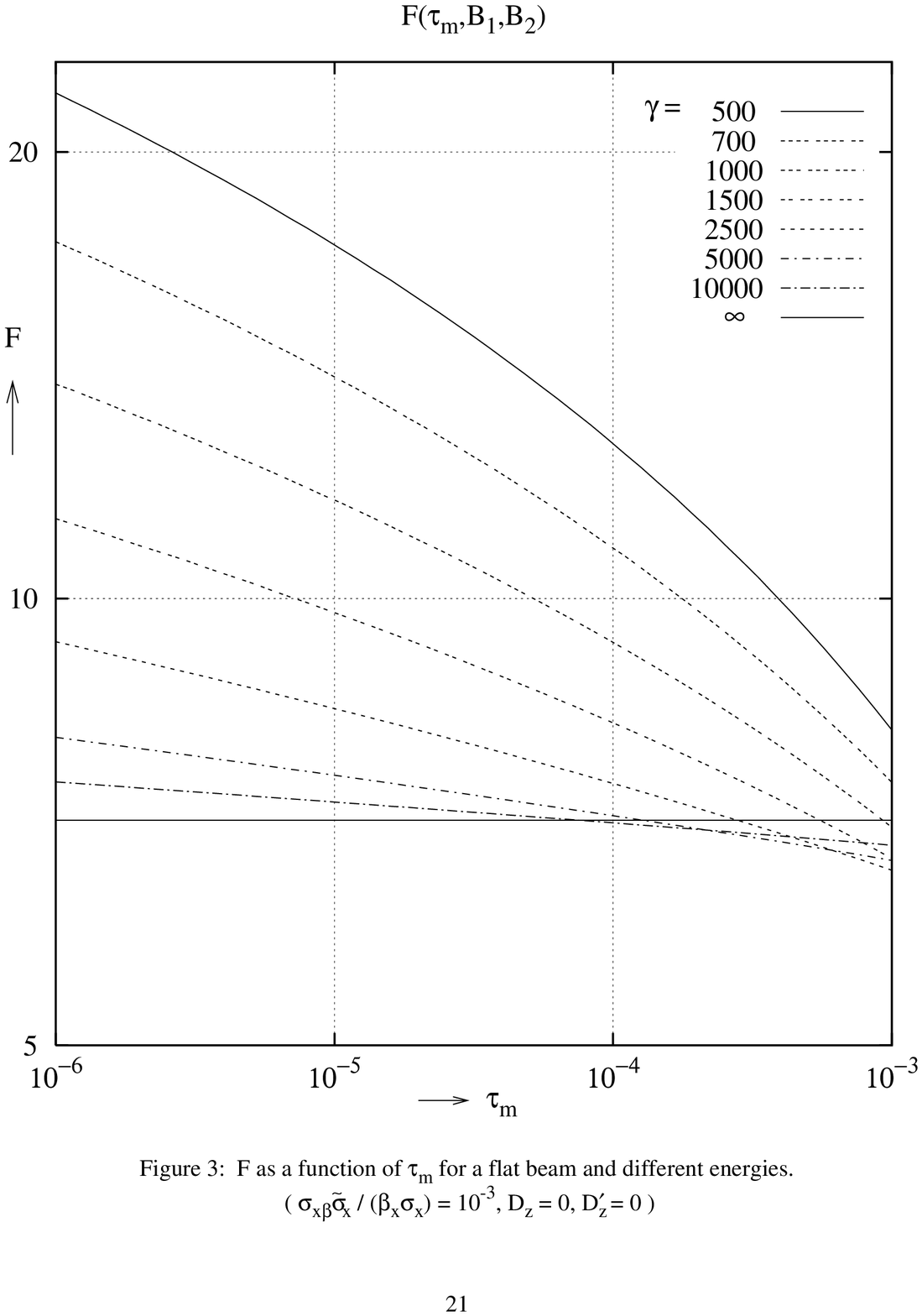,height=7.5in,width=7.in}
\end{figure}

\newpage

\pagestyle{empty}
\begin{figure}[p]
\psfig{figure=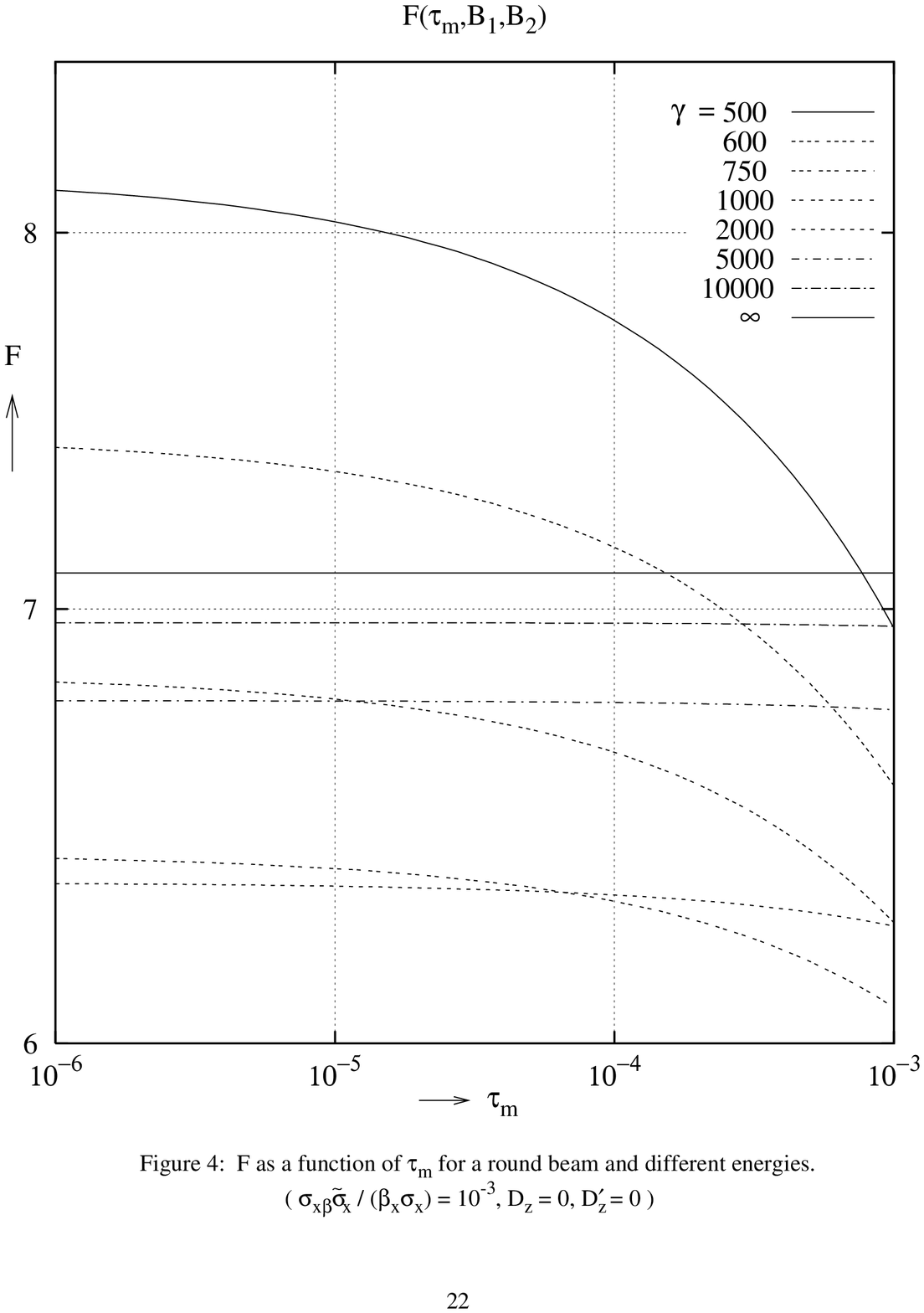,height=7.5in,width=7.in}
\end{figure}

\enddocument